\documentclass{Interspeech2024}

\usepackage{multirow}
\usepackage{array} 
\usepackage[subrefformat=parens]{subcaption}
\usepackage{comment}
\usepackage{cite}

\interspeechcameraready

\title{Lightweight Zero-shot Text-to-Speech with Mixture of Adapters}

\name[]{Kenichi}{Fujita}
\name[]{Takanori}{Ashihara}
\name[]{Marc}{Delcroix}
\name[]{Yusuke}{Ijima}

\address{
  NTT Corporation, Japan
}
\email{kenichi.fujita@ntt.com}


\begin{document}

\maketitle

\begin{abstract}

    The advancements in zero-shot text-to-speech (TTS) methods, based on large-scale models, have demonstrated high fidelity in reproducing speaker characteristics. However, these models are too large for practical daily use. We propose a lightweight zero-shot TTS method using a mixture of adapters (MoA). Our proposed method incorporates MoA modules into the decoder and the variance adapter of a non-autoregressive TTS model. These modules enhance the ability to adapt a wide variety of speakers in a zero-shot manner by selecting appropriate adapters associated with speaker characteristics on the basis of speaker embeddings. Our method achieves high-quality speech synthesis with minimal additional parameters. Through objective and subjective evaluations, we confirmed that our method achieves better performance than the baseline with less than 40\% of parameters at 1.9 times faster inference speed. 
\end{abstract}

\vspace{-6pt}
\section{Introduction}
\vspace{-4pt}
Advancements in text-to-speech (TTS) synthesis have enabled high-quality natural-sounding speech generation by leveraging large amounts of single and multi-speaker speech data~\cite{shen2018natural,ren2020fastspeech, chien2021investigating}. This has facilitated the development of TTS for diverse speakers using a minimal amount of target-speaker utterances. This capability extends to zero-shot scenarios in which the acoustic model can adapt without retraining~\cite{cooper2020zero,wang2023neural,fujita2023zeroshot}. Following the successes of large-scale language models~\cite{NEURIPS2020_1457c0d6} in zero-shot and few-shot adaptation, zero-shot TTS methods have achieved high fidelity by using large-scale models~\cite{wang2023neural,jiang2023megatts}. However, these methods are not well-suited for daily applications such as conversation robots, virtual assistants, and personalized TTS services, as they can not run on edge devices such as smartphones due to their substantial parameter sizes. While there is a demand, synthesizing high-quality speech under a zero-shot condition with a lightweight TTS model remains an ongoing challenge. Achieving this presents a significant challenge, as it requires maintaining the modeling capability to capture the diverse characteristics of thousands or more speakers with a model that has limited expressiveness due to its constrained number of parameters. 

To meet the demand for lightweight TTS, various methods have emerged, including autoregressive~\cite{kang2021fast,vainer20_interspeech}, non-autoregressive~\cite{NEURIPS2021_748d6b6e,Li2021LightTTS}, and diffusion-based~\cite{chen23_lightgrad} methods. However, none of these methods meets the requirement for lightweight zero-shot TTS. While PortaSpeech~\cite{NEURIPS2021_748d6b6e} and LightGrad~\cite{chen23_lightgrad} are lightweight, they are designed for single-speaker TTS, which only requires lower modeling ability. Light-TTS~\cite{Li2021LightTTS} is a multi-speaker TTS method but is only trained with about a few hundred speakers, which is not enough to achieve natural zero-shot TTS, as suggested with models trained with a few thousand speakers~\cite{cooper2020zero,wang2023neural,fujita2023zeroshot}. 

Designing models that are both expressive and parameter-efficient is a common challenge across fields such as classical pattern recognition to large models in natural language processing (NLP) and computer vision. One representative approach to address this is mixture of experts (MoE), with which multiple parallel expert modules are used, and one or more are selectively activated~\cite{Jacobs1991Adaptive,haykin1998neural,tresp2018committee,shazeer2017,Delcroix2018adaptive,NEURIPS2021_48237d9f, William2022Switch}. Determining the weights on these experts, i.e., selecting the appropriate expert using an input sequence or related information, enables the model to effectively handle diverse tasks. Consequently, MoE enhances model capacity while roughly maintaining training and inference efficiency by minimal additional parameters. Therefore, we embrace the concept of mixture of adapters (MoA)~\cite{chronopoulou-etal-2023-adaptersoup,wang-etal-2022-adamix}, an MoE variant primarily used in NLP. ADAPTERMIX~\cite{mehrish23_interspeech} leverages MoA for speaker adaptation in TTS, focusing on using stronger adapters to capture fine-grained speaker characteristics within the training data, i.e., few-shot adaptation, without delving into its potential for zero-shot TTS. Therefore, the strategy for adapter selection does not explicitly use speaker information.

We propose a lightweight zero-shot TTS method that is based on the concept of MoA. The key idea involves MoA gated with speaker embeddings. The proposed method is able to change the network configuration depending on the speaker characteristics. Therefore, it enables, at inference time, the arrangement of an efficient model adapted to the speaker using a small amount of speaker data. Because the model with MoA is trained on a large training dataset containing many speakers, the proposed method can cover various speaker characteristics at inference. To evaluate the proposed method, we conducted objective and subjective evaluations while varying the model size. The experimental results indicate that the insertion of MoA modules significantly enhances the lightweight model with minimal additional parameters and that the proposed method achieves better modeling ability than baselines by only using less than 40\% of parameters at 1.9 times faster inference speed. Audio samples are available on our demo page\footnote{https://ntt-hilab-gensp.github.io/is2024lightweightTTS/}.

\begin{figure*}[tb]
  \centering
  \begin{minipage}[b]{0.35\linewidth}
    \begin{minipage}[b]{1.0\linewidth}
      \centering  
      \includegraphics[width=0.85\linewidth]{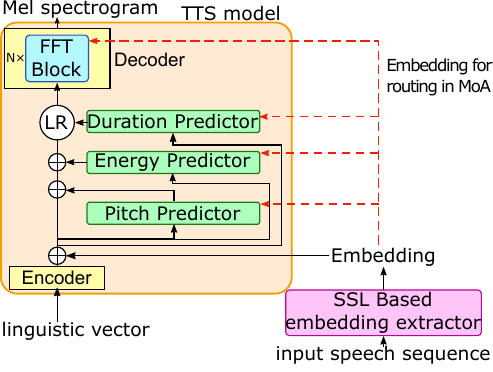}
      \vspace{-1mm}
      \subcaption{Overview of proposed method}
      \label{fig:overview_sep}
    \end{minipage}
  \end{minipage}
  \begin{minipage}[b]{0.17\linewidth}
    \begin{minipage}[b]{1.0\linewidth}
      \centering  
      \includegraphics[width=0.7\linewidth]{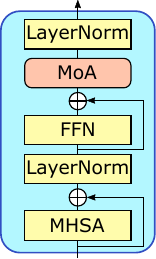}
      \vspace{-1mm}
      \subcaption{FFT with MoA}
      \label{fig:adap_trans}
    \end{minipage}
  \end{minipage}
  \begin{minipage}[b]{0.17\linewidth}
    \begin{minipage}[b]{1.0\linewidth}
      \centering  
      \includegraphics[width=0.7\linewidth]{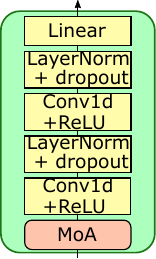}
      \vspace{-1mm}
      \subcaption{Predictor with MoA}
      \label{fig:adap_predictor}
    \end{minipage}
  \end{minipage}
  \begin{minipage}[b]{0.29\linewidth}
    \begin{minipage}[b]{1.0\linewidth}
      \centering  
      \includegraphics[width=1.0\linewidth]{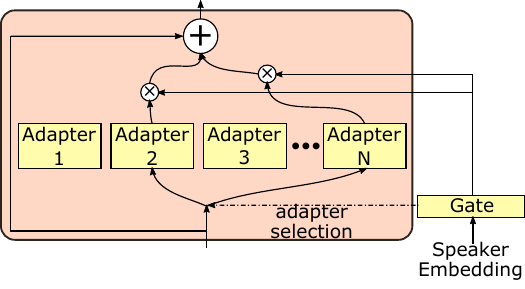}
      \vspace{-1mm}
      \subcaption{Overview of MoA module}
      \label{fig:MoA}
      \label{fig:moa}
    \end{minipage}
  \end{minipage}  
  \vspace{-3.5mm} 
  \caption{Overview of proposed method. MoA are introduced into Transformer decoder and predictors for pitch, energy, and duration.}
  \label{fig:overview_adapter}
  \vspace{-7mm}  
\end{figure*}

\vspace{-6pt}
\section{Proposed method}
\vspace{-4pt}
With our method, we expand the TTS model with MoA modules. Zero-shot TTS models typically comprise three main components: the TTS model with encoder and decoder, speaker-embedding extractor, and vocoder. Considering the utility of a TTS method, speaker embeddings can be extracted and stored before speech generation, enabling extraction on powerful computational devices or servers and eliminating the need for lightweight speaker embedding extractors. Therefore, we use a TTS method using a speaker-extraction method that is based on a self-supervised learning (SSL) speech model~\cite{fujita2023zeroshot}, which has demonstrated superior speech quality compared with conventional speaker recognition-based methods, e.g., d-vector~\cite{heigold2016end,doddipatla2017speaker} and x-vector~\cite{cooper2020zero, snyder2018x}. For the vocoder, lightweight methods have recently emerged, including those based on inverse short-time Fourier transform~\cite{kaneko2022istftnet,siuzdak2023vocos}, and some have zero-shot ability~\cite{siuzdak2023vocos}. These recent vocoders can achieve a low real-time factor (RTF), even lower than that of the TTS models. Consequently, it is important to develop lightweight TTS models for decreasing inference time of the entire system. Therefore, we concentrate on a lightweight TTS model, excluding the speaker embedding extractor and vocoder from the discussion. We now briefly introduce the backbone TTS model and MoA modules. 

\vspace{-4pt}
\subsection{Backbone SSL-based TTS model}
\label{sec:ssl_model}
\vspace{-4pt}

The proposed method uses the SSL-based embedding extractor to process input speech sequences. This extractor consists of an SSL model followed by an embedding module, which converts the speech representations from the SSL model into a fixed-length vector, i.e., a speaker embedding. The embedding module comprises three components: weighted-sum, bidirectional GRU, and attention. In the weighted-sum component, the speech representations from each layer of the SSL model are weighted using learnable weights then summed, following the same approach described in a previous paper~\cite{chen22g_interspeech}. Subsequently, the bidirectional GRU processes these summed representations, and their hidden states are further aggregated through an attention layer~\cite{bhattacharya2017deep, ando2018soft}. Finally, the obtained speaker embeddings are fed into the TTS model. As both the TTS model and embedding module are jointly trained, suitable speaker embeddings for the TTS model are obtained from the embedding module. During inference, we can use the embedding extractor separately from the TTS model and compute the speaker embedding in advance, similar to d-vector and x-vector. 

\vspace{-4pt}
\subsection{Speaker embedding based MoA}
\vspace{-4pt}
Figures~\ref{fig:adap_trans} and~\ref{fig:adap_predictor} respectively show an overview of the feed-forward Transformer (FFT) block of the decoder and the predictors (i.e., pitch, energy, and duration predictors) with an inserted MoA module (Fig.~\ref{fig:MoA}). This module comprises $N$ lightweight bottleneck adapters, each consisting of two feed-forward layers with layer normalization~\cite{chen22_exploring}, and a trainable gating network that determine the weights on the adapters using speaker embeddings. All components of the networks are jointly trained with the backbone TTS model. 

The MoA module is expressed as follows:
\vspace{-2pt}
\begin{equation}
  \text{MoA}(\mathbf{x}, \mathbf{x_e}) = \mathbf{x} + \sum_{i=1}^{N} g_i(\mathbf{x_e}) \cdot \text{Adapter}_i(\mathbf{x})
\end{equation}
where $\mathbf{x} \in \mathbb{R}^D$ is the input, $\mathbf{x_e} \in \mathbb{R}^{D_{\text{emb}}}$ is the speaker embedding, $\text{Adapter}_i: \mathbb{R}^D \mapsto \mathbb{R}^D$ represents an adapter from a set $\{\text{Adapter}_i(\mathbf{x})\}_{i=1}^{N}$ of $N$ adapters, and $g_i: \mathbb{R}^{D_{\text{emb}}} \mapsto \mathbb{R}^N$ is the trainable gating network parameterized by neural networks. We investigated two approaches for MoA. First, a \emph{dense} MoA, where the summation is executed on all the adapters $N$. Then a \emph{sparse} version, with which we only keep the top-$k$ $g_i$ weights and set the other weights to zero\footnote{We re-normalize the sparse weights to sum to one.}. The sparse version enables high representation power by having a large number of adapters during training while reducing the inference time. To encourage balanced load across weights for adapters, our method trained models with multi-task objectives, where the loss consists of the standard mean square error (MSE) losses and an additional auxiliary loss i.e., importance loss ($L_{importance}$)~\cite{shazeer2017}, defined as
\vspace{-2pt}
\begin{align}
  L_{importance}(\mathbf{X}) &= \genfrac{(}{)}{1pt}{}{\sigma(\text{Importance}(\mathbf{X}))}{\mu(\text{Importance}(\mathbf{X}))}^2 \\
  \text{Importance}(\mathbf{X}) &= \sum_{\mathbf{x_e} \in \mathbf{X}} g_i(\mathbf{x_e})
\end{align}
\vspace{-2pt}
where $\mathbf{X} \in \mathbb{R}^{n \times D}$ is the batch of speaker embeddings, and $\mu$ and $\sigma$ are average and standard deviation of the sequence.

\begin{figure*}[tb]
  \begin{minipage}[b]{0.32\linewidth}
    \centering
    \includegraphics[width=1.0\linewidth]
    {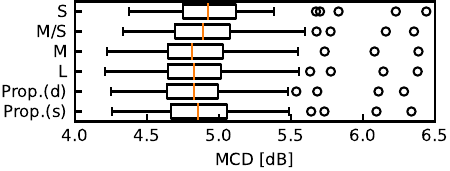}
    \vspace*{-5mm}  
    \subcaption{MCD (all)}\label{Fig_MCD}
   \end{minipage}
   \begin{minipage}[b]{0.32\linewidth}
    \centering
    \includegraphics[width=1.0\linewidth]
    {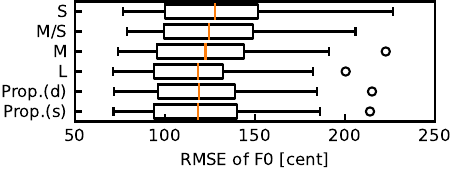}
    \vspace*{-5mm}  
    \subcaption{F0 RMSE (all)}\label{Fig_F0}
   \end{minipage}
   \begin{minipage}[b]{0.32\linewidth}
    \centering
    \includegraphics[width=1.0\linewidth]
    {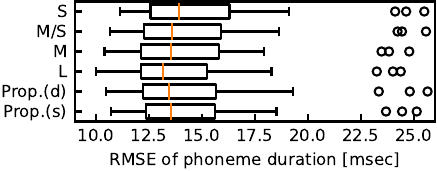}
    \vspace*{-5mm}  
    \subcaption{Dur. RMSE (all)}\label{Fig_dur}
   \end{minipage}\\
   \begin{minipage}[b]{0.32\linewidth}
    \centering
    \includegraphics[width=1.0\linewidth]
    {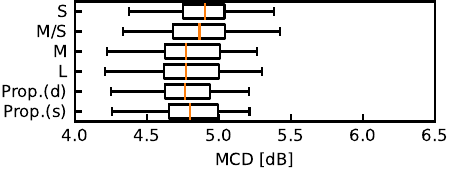}
    \vspace*{-5mm}  
    \subcaption{MCD (nonpro.)}\label{Fig_MCD_nonpro}
   \end{minipage}
   \begin{minipage}[b]{0.32\linewidth}
    \centering
    \includegraphics[width=1.0\linewidth]
    {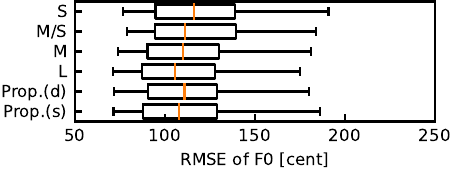}
    \vspace*{-5mm}  
    \subcaption{F0 RMSE (nonpro.)}\label{Fig_F0_nonpro}
   \end{minipage}
   \begin{minipage}[b]{0.32\linewidth}
    \centering
    \includegraphics[width=1.0\linewidth]
    {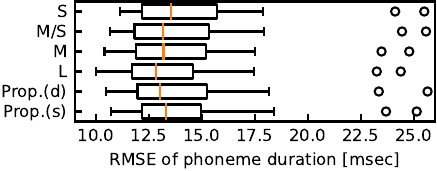}
    \vspace*{-5mm}  
    \subcaption{Dur. RMSE (nonpro)}\label{Fig_dur_nonpro}
   \end{minipage}\\
   \begin{minipage}[b]{0.32\linewidth}
    \centering
    \includegraphics[width=1.0\linewidth]
    {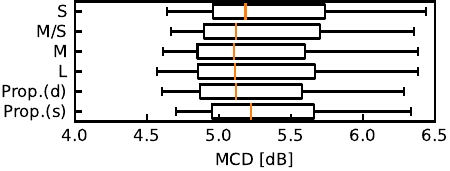}
    \vspace*{-5mm}  
    \subcaption{MCD (pro.)}\label{Fig_MCD_pro}
   \end{minipage}
   \begin{minipage}[b]{0.32\linewidth}
    \centering
    \includegraphics[width=1.0\linewidth]
    {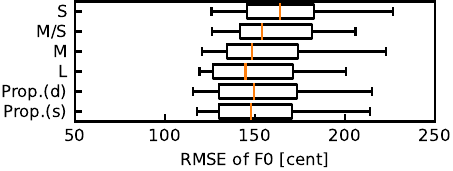}
    \vspace*{-5mm}  
    \subcaption{F0 RMSE (pro.)}\label{Fig_F0_pro}
   \end{minipage}
   \begin{minipage}[b]{0.32\linewidth}
    \centering
    \includegraphics[width=1.0\linewidth]
    {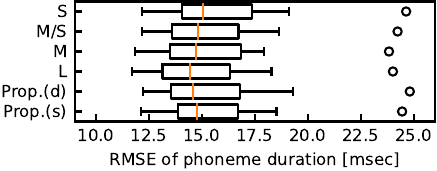}
    \vspace*{-5mm}  
    \subcaption{Dur. RMSE (pro.)}\label{Fig_dur_pro}
   \end{minipage}\\
     \vspace*{-7mm}  
  \caption{Results of objective evaluations (lower the better). Dur. represents phoneme duration and pro. and nonpro. represent professional and non-professional speakers.}\label{obeval}
  \vspace*{-6mm}  
\end{figure*}

\vspace{-6pt}
\section{Experimental Setup}
\vspace{-4pt}
\subsection{Dataset}
\vspace{-4pt}
For training TTS models, we used an in-house 960-hour Japanese speech database which includes 5,362 speakers: 3,242 female and 2,120 male. This database includes several speaker types, such as professional speakers, i.e., newscasters, narrators, and voice actors, as well as non-professional speakers. Among the female and male speakers, 160 and 92 are professionals, respectively. The database was split into three parts: 303,406 utterances by 5,296 speakers for training, 6,807 by 50 speakers for validation (26 females and 25 males, including 6 and 5 professional speakers, respectively), and 6,809 by 64 speakers for testing (35 females and 30 males, including 9 and 5 professional speakers, respectively). The sampling frequency of the speech was \SI{22}{\kHz}.

\vspace{-4pt}
\subsection{Training conditions}
\vspace{-4pt}
The TTS model was FastSpeech2, as implemented in a previous study~\cite{chien2021investigating}, featuring four-layer Transformers for the encoder and six-layer Transformers for the decoder. To confirm the effectiveness of MoA insertion and compare it with models having larger parameters, we trained four models: \textit{Small}~(\textit{S}), \textit{Medium Small}~(\textit{M/S}), \textit{Medium}~(\textit{M}), and \textit{Large}~(\textit{L}), with 14, 19, 42, and 151M parameters, respectively, obtained by varying the hidden dimensions and filter size of the decoder and encoder and the filter size of the predictors. Table~\ref{table:hyper_param} outlines their hyperparameters. 

\begin{table}[tb]
  \caption{Model hyper-parameters and their RTF. RTF was calculated on single thread CPU (Intel(R) Core(TM) i9-10940X CPU @ 3.30GHz). \textit{Prop.} represents \textit{Proposed}. }
  \vspace*{-7mm}  
  \label{table:hyper_param}
  \begin{center}
  \scalebox{0.9}{
  \begin{tabular}{wc{24mm}|wc{5.5mm}wc{5.5mm}wc{5.5mm}wc{5.5mm}wc{6mm}wc{6mm}}
  \hline
  \noalign{\vskip.5mm}
    Model& \textit{S}& \textit{M/S} & \textit{M} & \textit{L} &\textit{Prop.(d)}&\textit{Prop.(s)}\\ 
  \hline
  Params & 14M &19M & 42M & 151M &15M& 16M\\
  Params (inference)  & 14M &19M &42M &151M &15M&15M\\  
  Enc./Dec. dimension & 128 &160 & 256 & 512 &128&128\\
  Enc./Dec. filter size & 256 &320 & 512 & 1024 &256&256\\
  Predictor's filter size & 256 &320 & 512 & 1024 &256&256\\
  RTF (inference) & 0.0127 &0.0167 & 0.0286 & 0.0804 &0.0148& 0.0148\\

  \noalign{\vskip.5mm}
  \hline
  \end{tabular}%
  }
  \end{center}
 \vspace*{-10mm}
\end{table}

MoA modules were inserted into the \textit{Small} model. To confirm the advantage of sparse gating, we conducted experiments with two types of MoA: sparse~(\textit{Proposed(s)}) and dense~(\textit{Proposed(d)}). In the former, there were 8 adapters ($N$=8), and the $k$ in top-$k$ sampling was set to 3, while in the latter, $N$ and $k$ were both set to 3, i.e., without top-$k$ sampling. Since both types use three adapters during inference, their computational cost at inference is identical. The bottleneck size of the adapters was 96. As shown in Table~\ref{table:hyper_param}, smaller parameter models achieved higher inference speed, and the additional MoA modules did not largely affect speed. 

The input and target sequences were a 303-dimensional linguistic vector and 80-dimensional mel-spectrograms with a \SI{10.0}{\ms} frame shift. We used the publicly available HuBERT {\sc BASE}~\cite{hsu2021hubert} trained with ReazonSpeech~\cite{fujimotoreazonspeech}\footnote{https://huggingface.co/rinna/japanese-hubert-base} as the SSL model. HuBERT processed the \SI{16}{kHz} raw audio input sequence into 768-dimensional sequences, and the embedding modules converted them into fixed-length vectors with the same size as the decoder dimension. For waveform generation, we used HiFi-GAN~\cite{NEURIPS2020_c5d73680}. Fine-tuning HiFi-GAN for each TTS model could improve naturalness and similarity, but for fair performance comparison between TTS models, we used HiFi-GAN without fine-tuning. While lighter vocoders are available, their usage is beyond the scope of our study.

To ensure stable MoA training, we initially pre-trained the whole backbone TTS model including the SSL-based embedding extractor with training data for 400K steps using the Adam optimizer~\cite{kingma-Adam}, following the same learning-rate schedule as Vaswani et al.~\cite{NIPS2017_3f5ee243}. Subsequently, adapters were introduced into the variance adapter (i.e., pitch, energy, and duration predictors), and decoder. The entire model including the SSL-based embedding extractor was then trained with the same training data for an additional 400K steps. Through the training process, HuBERT remained frozen. For equitable comparison, all other models trained without MoA modules underwent a total training duration of 800K steps.

\begin{figure*}[tb]
  \begin{minipage}[b]{0.22\linewidth}
   \centering
   \includegraphics[width=0.95\linewidth]
   {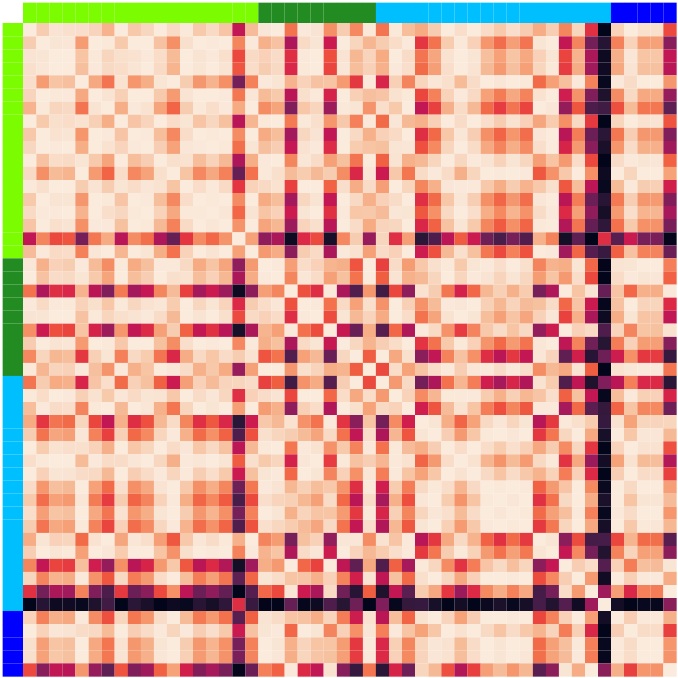}
   \vspace*{-1mm}
   \subcaption{1st layer (\textit{Prop.(d)})}\label{fig:weights_dense_1}
  \end{minipage}
  \begin{minipage}[b]{0.22\linewidth}
    \centering
    \includegraphics[width=0.95\linewidth]
    {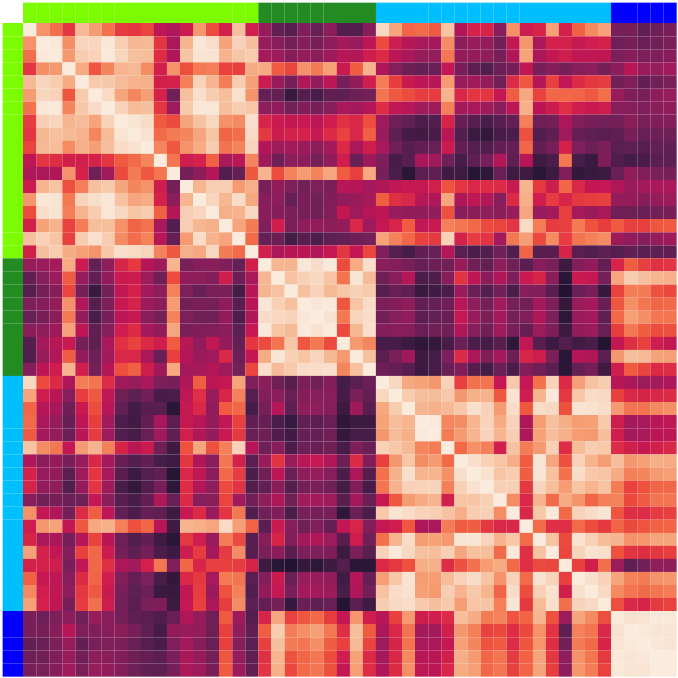}
    \vspace*{-1mm}
    \subcaption{1st layer (\textit{Prop.(s)})}\label{fig:weights_sprse_1}
   \end{minipage}
   \begin{minipage}[b]{0.22\linewidth}
    \centering
    \includegraphics[width=0.95\linewidth]
    {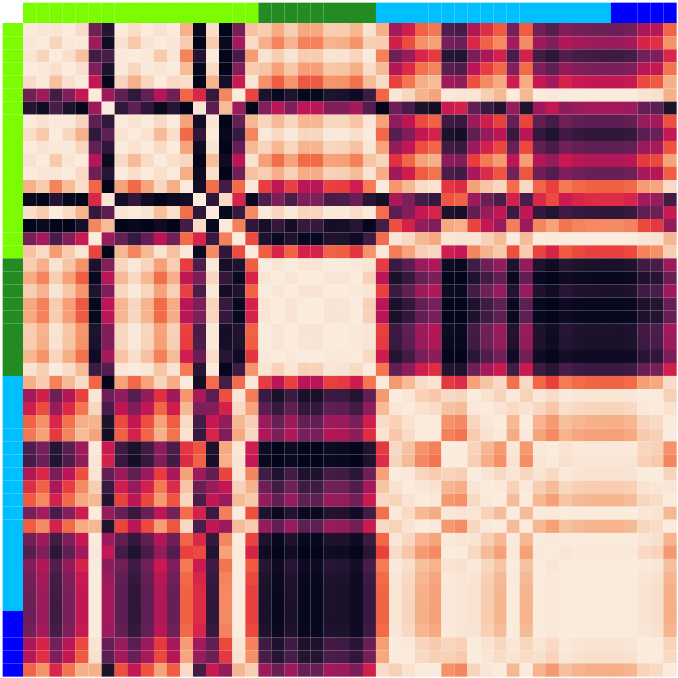}
    \vspace*{-1mm}
    \subcaption{3rd layer (\textit{Prop.(d)})}\label{fig:weights_dense_3}
   \end{minipage}
   \begin{minipage}[b]{0.22\linewidth}
    \centering
    \includegraphics[width=0.95\linewidth]
    {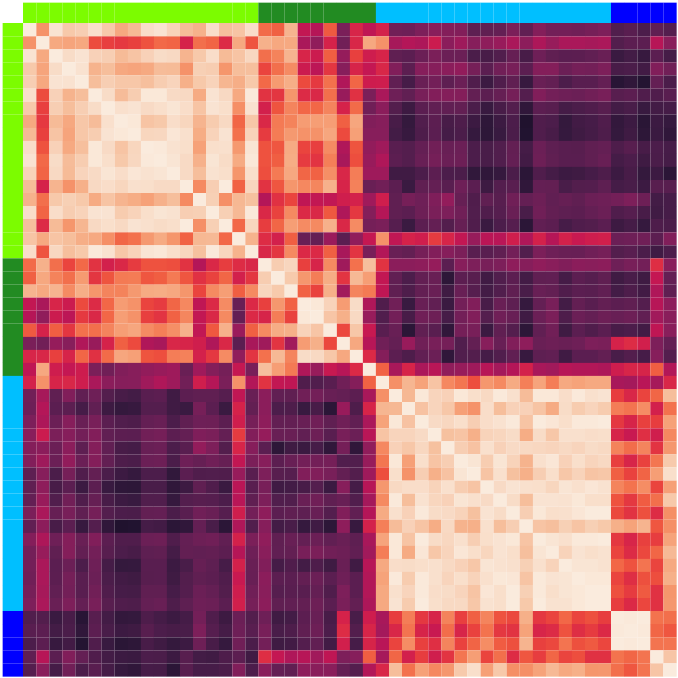}
    \vspace*{-1mm}
    \subcaption{3rd layer (\textit{Prop.(s)})}\label{fig:weights_sparse_3}
   \end{minipage}
   \begin{minipage}[b]{0.1\linewidth}
    \centering
    \includegraphics[width=1.0\linewidth]
    {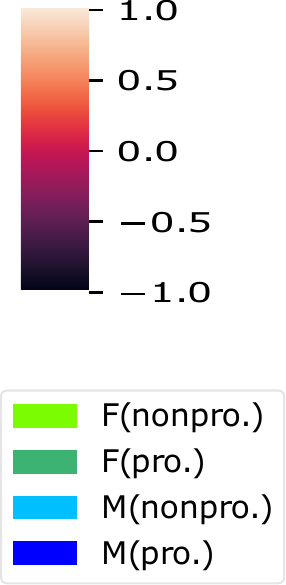}
    \vspace*{0mm}
   \end{minipage}
   \vspace*{-7mm}    
  \caption{Correlation coefficient of weights for adapters among test speakers. Colored group labels show the characteristics of each speaker (Female and Male, professional and non-professional speakers).}\label{fig:adapter_weights}
  \vspace*{-6.5mm}    
  \end{figure*}

\vspace*{-6pt} 
\section{Results}
\vspace*{-4pt} 
\subsection{Objective evaluations}
\vspace*{-4pt} 
We first conducted an objective evaluation to evaluate the performance of the proposed method under a \textit{data-parallel} condition, where the text to be synthesized matches the text of the reference speech, i.e., the input speech sequence to the embedding extractor. Past research~\cite{fujita2023zeroshot} indicates that objective metrics may not accurately reflect the reproduction ability under a \textit{non-parallel} condition due to intra-speaker variation, i.e., slight inconsistencies within utterances from the same speaker, attributable to the high reproduction quality of SSL-based TTS models for reference speech. The evaluation metrics included mel-cepstral distortion (MCD), and root MSE (RMSE) of logarithmic F0 and phoneme durations. To calculate MCD and F0 RMSE between the generated and test speech with the same time alignment, without using dynamic time warping, we generated a log mel-spectrogram using the original phoneme durations extracted from the test speech. The RMSE of the phoneme durations was computed by comparing the original phoneme durations of the test speech with those predicted by the duration predictor. 

Figure~\ref{obeval} shows the objective evaluation results. We calculated the average of each metric per speaker, and the figures illustrate their distributions. The third quartile is the point, as it serves as a good indicator of modeling ability, reflecting the performance in reproducing challenging speakers. A comparison across models with varying parameter sizes, i.e., \textit{L}, \textit{M}, \textit{M/S}, and \textit{S}, revealed that performance degraded as the number of parameters decreased. This suggests that simply reducing the dimensions in the TTS model could compromise speech generation performance, as smaller models have lower modeling ability.

\textit{Proposed(d)} and \textit{Proposed(s)} exhibited superior performance compared with \textit{S} and \textit{M/S} across all metrics, indicating that inserting the MoA module improves performance and the improvement is larger than simply adding parameters. Even when compared with \textit{M}, \textit{Proposed(d)} and \textit{Proposed(s)} still demonstrated better or comparable performance despite having less than 40\% of the parameters, enabling 1.9 times faster inference speed (RTF 0.0148 vs 0.0286).

Comparing the results from non-professional and professional speakers, the objective metrics of the former were better than those for the latter. This may be because professional speakers generally have more dynamic speaking styles, which are more challenging to reproduce with a TTS system. Additionally, the number of professional speakers in the training data was much smaller compared with that of non-professional speakers. However, \textit{Proposed(d)} and \textit{Proposed(s)} even performed better or comparably to \textit{M}.

\begin{table}[tb]
  \caption{Preference scores from AB test in naturalness from subjective evaluation (A-N/P-B). Bold score indicates preferred method has p-value less than 0.05 in chi-square test with BH correction.}
  \vspace*{-7mm}
  \label{table:AB}
  \begin{center}
  \scalebox{0.9}{    
  \begin{tabular}{wc{10mm}wc{21mm}wc{21mm}wc{21mm}}
  \hline
  \noalign{\vskip.5mm}
   & \textit{Prop.(s)} vs \textit{S} & \textit{Prop.(s)} vs \textit{M}&  \textit{Prop.(s)} vs \textit{Prop.(d)}\\
  \hline
  \textit{All} & \textbf{0.52} - 0.29 - 0.19& \textbf{0.36} - 0.36 - 0.28& \textbf{0.33} - 0.47 - 0.20\\
  \textit{nonpro.} & \textbf{0.42} - 0.34 - 0.24& 0.33 - 0.34 - 0.33& \textbf{0.30} - 0.48 - 0.22\\
  \textit{pro.} & \textbf{0.62} - 0.24 - 0.14& \textbf{0.38} - 0.39 - 0.23 & \textbf{0.37} - 0.45 - 0.18\\
  \noalign{\vskip.5mm}
  \hline
  \end{tabular}%
  }
  \end{center}
 \vspace*{-8mm}    
\end{table}

\begin{table}[tb]
  \caption{Preference scores from XAB test in similarity from subjective evaluation (A-N/P-B). Bold score indicates preferred method has p-value less than 0.05 in chi-square test with BH correction.}
  \vspace*{-7mm}  
  \label{table:XAB}
  \begin{center}
  \scalebox{0.9}{    
  \begin{tabular}{wc{10mm}wc{21mm}wc{21mm}wc{21mm}}
  \hline
  \noalign{\vskip.5mm}
   & \textit{Prop.(s)} vs \textit{S} & \textit{Prop.(s)} vs \textit{M}&  \textit{Prop.(s)} vs \textit{Prop.(d)}\\
  \hline
  \textit{All} & \textbf{0.46} - 0.27 - 0.27& \textbf{0.36} - 0.36 - 0.28& \textbf{0.32} - 0.46 - 0.22\\
  \textit{nonpro.} & 0.34 - 0.38 - 0.28& 0.36 - 0.35 - 0.29& 0.27 - 0.50 - 0.23\\
  \textit{pro.} & \textbf{0.56} - 0.16 - 0.28& \textbf{0.37} - 0.37 - 0.26 & \textbf{0.38} - 0.41 - 0.21\\
  \noalign{\vskip.5mm}
  \hline
  \end{tabular}%
  }
  \end{center}
  \vspace*{-10mm}    
\end{table}

\vspace*{-4pt}
\subsection{Subjective evaluation}
\label{sec:subjective_eval}
\vspace*{-4pt}
We conducted a subjective evaluation on naturalness and similarity with AB and XAB listening tests, respectively. All permutations of synthetic speech pairs were presented in two orders (AB and BA, and XAB and XBA) to eliminate bias in the order of stimuli. The experiments were conducted involving 16 participants. In the AB test, each participant was presented synthesized speech samples and asked to indicate the one with better naturalness or show no preference (N/P). In the XAB test, each participant determined which one was more similar to the ground-truth recorded speech from the target speaker, or if they were equally similar. On the basis of the results of the objective evaluation, we compared \textit{Proposed(s)} with: \textit{S}, \textit{M}, and \textit{Proposed(d)}. Each model synthesized ten sentences under the \textit{non-parallel} condition from each of the four speakers in the test data, i.e., professional and non-professional males and females, respectively. 

Table~\ref{table:AB} lists the preference scores for each model. \textit{Proposed(s)} exhibited significantly better naturalness than \textit{S}, \textit{M}, and \textit{Proposed(d)}. \textit{Proposed(s)} also exhibited significantly better similarity than \textit{S}, \textit{M} and \textit{Proposed(d)}, as shown in Table~\ref{table:XAB}. This suggests that using MoA improves performance, with the improvement even reaching to the models with more than double the parameters, i.e., the TTS model with MoA performed better with less than 40\% of parameters at 1.9 times faster inference speed, as shown in Table~\ref{table:hyper_param}. The difference between \textit{Proposed(d)} and \textit{Proposed(s)} indicates that sparse gating and having a larger number of adapters could enhance performance while maintaining inference computational cost. The results further indicate the significant strength of \textit{Proposed(s)} across professional speakers compared with the other models.

\vspace{-4pt}
\subsection{Weight analysis at gating function}
\vspace{-4pt}
To analyze the improved performance with MoA, we examined the weights on experts at each layer of the decoder. We computed the correlation coefficient of weights on adapters among the speakers in the test data. The utterance of each speaker was randomly selected, ensuring that the content differed among all the speakers. 

Figure~\ref{fig:adapter_weights} displays the heatmaps of the correlation coefficients from the 1st and 3rd layers of the decoder. Figures~\ref{fig:weights_dense_1} and \ref{fig:weights_dense_3} are those from \textit{Proposed(d)}, while Figs.~\ref{fig:weights_sprse_1} and \ref{fig:weights_sparse_3} are those from \textit{Proposed(s)}. \textit{Proposed(s)} revealed higher correlation coefficients among speakers with similar characteristics and lower among others. This indicates that characteristic-specific experts were obtained, i.e., the specialized adapters were appropriately selected from adapter pools. \textit{Proposed(d)} also seemed to reflect characteristics such as female and male. However, it appeared to fail to differentiate between professional and non-professional speakers. Overall, \textit{Proposed(s)} would obtain more expressive experts to handle a more diverse range of speakers. The deeper the layer, the more detailed heatmaps seems to appear, suggesting a progression from coarse to refined features through the decoder. These suggest that having a larger number of adapters and selecting them through sparse gating may lead to better speaker representation, covering a wider range of speakers with diverse expert adapters.

\vspace{-6pt}
\section{Conclusion}
\vspace{-4pt}
We proposed a lightweight zero-shot TTS method using MoA. Objective and subjective evaluations showed that the proposed method achieves higher naturalness and similarity with minimal additional parameters and computational cost. Furthermore, the method performs better than the baseline model with less than 40\% of parameters at 1.9 times faster inference speed. Future work includes applying the proposed method to larger TTS models including the different structure models such as VALL-E~\cite{wang2023neural} and confirming that it still contributes to quality improvements.

\clearpage 
\bibliographystyle{IEEEtran}
\bibliography{mybib,refs}

\begin{thebibliography}{10}
\providecommand{\url}[1]{#1}
\csname url@samestyle\endcsname
\providecommand{\newblock}{\relax}
\providecommand{\bibinfo}[2]{#2}
\providecommand{\BIBentrySTDinterwordspacing}{\spaceskip=0pt\relax}
\providecommand{\BIBentryALTinterwordstretchfactor}{4}
\providecommand{\BIBentryALTinterwordspacing}{\spaceskip=\fontdimen2\font plus
\BIBentryALTinterwordstretchfactor\fontdimen3\font minus \fontdimen4\font\relax}
\providecommand{\BIBforeignlanguage}[2]{{%
\expandafter\ifx\csname l@#1\endcsname\relax
\typeout{** WARNING: IEEEtran.bst: No hyphenation pattern has been}%
\typeout{** loaded for the language `#1'. Using the pattern for}%
\typeout{** the default language instead.}%
\else
\language=\csname l@#1\endcsname
\fi
#2}}
\providecommand{\BIBdecl}{\relax}
\BIBdecl

\bibitem{shen2018natural}
J.~Shen, R.~Pang, R.~J. Weiss, M.~Schuster, N.~Jaitly, Z.~Yang \emph{et~al.}, ``Natural {TTS} synthesis by conditioning {WaveNet} on mel spectrogram predictions,'' in \emph{ICASSP}, 2018, pp. 4779--4783.

\bibitem{ren2020fastspeech}
Y.~Ren, C.~Hu, X.~Tan, T.~Qin, S.~Zhao, Z.~Zhao, and T.-Y. Liu, ``{FastSpeech} 2: Fast and high-quality end-to-end text to speech,'' in \emph{ICLR}, 2020.

\bibitem{chien2021investigating}
C.-M. Chien, J.-H. Lin, C.-y. Huang, P.-c. Hsu, and H.-y. Lee, ``Investigating on incorporating pretrained and learnable speaker representations for multi-speaker multi-style text-to-speech,'' in \emph{ICASSP}, 2021, pp. 8588--8592.

\bibitem{cooper2020zero}
E.~Cooper, C.-I. Lai, Y.~Yasuda, F.~Fang, X.~Wang, N.~Chen, and J.~Yamagishi, ``Zero-shot multi-speaker text-to-speech with state-of-the-art neural speaker embeddings,'' in \emph{ICASSP}, 2020, pp. 6184--6188.

\bibitem{wang2023neural}
C.~Wang, S.~Chen, Y.~Wu, Z.~Zhang, L.~Zhou, S.~Liu \emph{et~al.}, ``Neural codec language models are zero-shot text to speech synthesizers,'' arXiv, 2023, arXiv:2301.02111.

\bibitem{fujita2023zeroshot}
K.~Fujita, T.~Ashihara, H.~Kanagawa, T.~Moriya, and Y.~Ijima, ``Zero-shot text-to-speech synthesis conditioned using self-supervised speech representation model,'' in \emph{ICASSP Workshops (ICASSPW)}, 2023, pp. 1--5.

\bibitem{NEURIPS2020_1457c0d6}
T.~Brown, B.~Mann, N.~Ryder, M.~Subbiah, J.~D. Kaplan, P.~Dhariwal \emph{et~al.}, ``Language models are few-shot learners,'' in \emph{NeurIPS}, vol.~33, 2020, pp. 1877--1901.

\bibitem{jiang2023megatts}
Z.~Jiang, Y.~Ren, Z.~Ye, J.~Liu, C.~Zhang, Q.~Yang, S.~Ji, R.~Huang, C.~Wang, X.~Yin, Z.~Ma, and Z.~Zhao, ``{Mega-TTS}: Zero-shot text-to-speech at scale with intrinsic inductive bias,'' arXiv, 2023, arXiv:2306.03509.

\bibitem{kang2021fast}
M.~Kang, J.~Lee, S.~Kim, and I.~Kim, ``{Fast DCTTS}: efficient deep convolutional text-to-speech,'' in \emph{ICASSP}, 2021, pp. 7043--7047.

\bibitem{vainer20_interspeech}
J.~Vainer and O.~Dusek, ``{SpeedySpeech}: Efficient neural speech synthesis,'' in \emph{Interspeech}, 2020, pp. 3575--3579.

\bibitem{NEURIPS2021_748d6b6e}
Y.~Ren, J.~Liu, and Z.~Zhao, ``{PortaSpeech}: Portable and high-quality generative text-to-speech,'' in \emph{NeurIPS}, vol.~34, 2021, pp. 13\,963--13\,974.

\bibitem{Li2021LightTTS}
S.~Li, B.~Ouyang, L.~Li, and Q.~Hong, ``{Light-TTS}: Lightweight multi-speaker multi-lingual text-to-speech,'' in \emph{ICASSP}, 2021, pp. 8383--8387.

\bibitem{chen23_lightgrad}
J.~Chen, X.~Song, Z.~Peng, B.~Zhang, F.~Pan, and Z.~Wu, ``{LightGrad}: Lightweight diffusion probabilistic model for text-to-speech,'' in \emph{ICASSP}, 2023, pp. 1--5.

\bibitem{Jacobs1991Adaptive}
R.~A. Jacobs, M.~I. Jordan, S.~J. Nowlan, and G.~E. Hinton, ``Adaptive mixtures of local experts,'' \emph{Neural Computation}, vol.~3, no.~1, pp. 79--87, 1991.

\bibitem{haykin1998neural}
S.~Haykin, \emph{Neural networks: a comprehensive foundation}.\hskip 1em plus 0.5em minus 0.4em\relax USA:Prentice Hall, 1994.

\bibitem{tresp2018committee}
V.~Tresp, ``Committee machines,'' in \emph{Handbook of neural network signal processing}, 2002.

\bibitem{shazeer2017}
N.~Shazeer, A.~Mirhoseini, K.~Maziarz, A.~Davis, Q.~Le, G.~Hinton, and J.~Dean, ``Outrageously large neural networks: The sparsely-gated mixture-of-experts layer,'' in \emph{ICLR}, 2017.

\bibitem{Delcroix2018adaptive}
M.~Delcroix, K.~Kinoshita, A.~Ogawa, C.~Huemmer, and T.~Nakatani, ``Context adaptive neural network based acoustic models for rapid adaptation,'' \emph{TASLP}, vol.~26, no.~5, pp. 895--908, 2018.

\bibitem{NEURIPS2021_48237d9f}
C.~Riquelme, J.~Puigcerver, B.~Mustafa, M.~Neumann, R.~Jenatton, A.~Susano~Pinto, D.~Keysers, and N.~Houlsby, ``Scaling vision with sparse mixture of experts,'' in \emph{NeurIPS}, vol.~34, 2021, pp. 8583--8595.

\bibitem{William2022Switch}
W.~Fedus, B.~Zoph, and N.~Shazeer, ``{Switch Transformers}: scaling to trillion parameter models with simple and efficient sparsity,'' \emph{JMLR}, vol.~23, no.~1, jan 2022.

\bibitem{chronopoulou-etal-2023-adaptersoup}
A.~Chronopoulou, M.~Peters, A.~Fraser, and J.~Dodge, ``{A}dapter{S}oup: Weight averaging to improve generalization of pretrained language models,'' in \emph{EACL}, 2023, pp. 2054--2063.

\bibitem{wang-etal-2022-adamix}
Y.~Wang, S.~Agarwal, S.~Mukherjee, X.~Liu, J.~Gao, A.~H. Awadallah, and J.~Gao, ``{A}da{M}ix: Mixture-of-adaptations for parameter-efficient model tuning,'' in \emph{EMNLP}, 2022, pp. 5744--5760.

\bibitem{mehrish23_interspeech}
A.~Mehrish, A.~{Ramesh Kashyap}, L.~Yingting, N.~Majumder, and S.~Poria, ``{ADAPTERMIX}: Exploring the efficacy of mixture of adapters for low-resource {TTS} adaptation,'' in \emph{Interspeech}, 2023, pp. 4284--4288.

\bibitem{heigold2016end}
G.~Heigold, I.~Moreno, S.~Bengio, and N.~Shazeer, ``End-to-end text-dependent speaker verification,'' in \emph{ICASSP}, 2016, pp. 5115--5119.

\bibitem{doddipatla2017speaker}
R.~Doddipatla, N.~Braunschweiler, and R.~Maia, ``Speaker adaptation in {DNN}-based speech synthesis using d-vectors.'' in \emph{Interspeech}, 2017, pp. 3404--3408.

\bibitem{snyder2018x}
D.~Snyder, D.~Garcia-Romero, G.~Sell, D.~Povey, and S.~Khudanpur, ``X-vectors: Robust {DNN} embeddings for speaker recognition,'' in \emph{ICASSP}, 2018, pp. 5329--5333.

\bibitem{kaneko2022istftnet}
T.~Kaneko, K.~Tanaka, H.~Kameoka, and S.~Seki, ``{iSTFTNet}: Fast and lightweight mel-spectrogram vocoder incorporating inverse short-time {Fourier} transform,'' in \emph{ICASSP}, 2022, pp. 6207--6211.

\bibitem{siuzdak2023vocos}
H.~Siuzdak, ``Vocos: Closing the gap between time-domain and {Fourier}-based neural vocoders for high-quality audio synthesis,'' in \emph{ICLR}, 2024.

\bibitem{chen22g_interspeech}
S.~Chen, Y.~Wu, C.~Wang, S.~Liu, Z.~Chen, P.~Wang \emph{et~al.}, ``Why does self-supervised learning for speech recognition benefit speaker recognition?'' in \emph{Interspeech}, 2022, pp. 3699--3703.

\bibitem{bhattacharya2017deep}
G.~Bhattacharya, M.~J. Alam, and P.~Kenny, ``Deep speaker embeddings for short-duration speaker verification.'' in \emph{Interspeech}, 2017, pp. 1517--1521.

\bibitem{ando2018soft}
A.~Ando, S.~Kobashikawa, H.~Kamiyama, R.~Masumura, Y.~Ijima, and Y.~Aono, ``Soft-target training with ambiguous emotional utterances for {DNN}-based speech emotion classification,'' in \emph{ICASSP}, 2018, pp. 4964--4968.

\bibitem{chen22_exploring}
Z.-C. Chen, C.-L. Fu, C.-Y. Liu, S.-W.~D. Li, and H.-y. Lee, ``Exploring efficient-tuning methods in self-supervised speech models,'' in \emph{SLT}, 2022, pp. 1120--1127.

\bibitem{hsu2021hubert}
W.-N. Hsu, B.~Bolte, Y.-H.~H. Tsai, K.~Lakhotia, R.~Salakhutdinov, and A.~Mohamed, ``{HuBERT}: Self-supervised speech representation learning by masked prediction of hidden units,'' \emph{TASLPA}, vol.~29, pp. 3451--3460, 2021.

\bibitem{fujimotoreazonspeech}
Y.~Yin, D.~Mori, and S.~Fujimoto, ``{ReazonSpeech}: A free and massive corpus for {Japanese} {ASR},'' 2023.

\bibitem{NEURIPS2020_c5d73680}
J.~Kong, J.~Kim, and J.~Bae, ``{HiFi-GAN}: Generative adversarial networks for efficient and high fidelity speech synthesis,'' in \emph{NeurIPS}, vol.~33, 2020, pp. 17\,022--17\,033.

\bibitem{kingma-Adam}
D.~P. Kingma and J.~Ba, ``Adam: A method for stochastic optimization,'' in \emph{ICLR}, 2015.

\bibitem{NIPS2017_3f5ee243}
A.~Vaswani, N.~Shazeer, N.~Parmar, J.~Uszkoreit, L.~Jones, A.~N. Gomez \emph{et~al.}, ``Attention is all you need,'' in \emph{NeurIPS}, vol.~30, 2017.

\end{thebibliography}

\end{document}